\documentclass[11pt]{article}
 
\usepackage{amssymb,fullpage,graphicx}
\usepackage{xcolor,soul}

\def\ld{\mathrm{ld}}
\def\domega{{\mathrm{d}\omega}}
\def\calP{\mathcal{P}}
\def\det{\mathrm{det}}
\def\bbR{\mathbb{R}}
\def\AGM{\mathrm{AGM}}
\def\AHM{\mathrm{AHM}}
\def\tr{\mathrm{tr}}
\def\farthest{\mathrm{farthest}}
\def\PD{\mathrm{PD}}
\def\mod{\mathrm{mod}}
\def\inner#1#2{\langle #1,#2\rangle}
\def\LEM{\mathrm{LE}}
\def\bbP{\mathbb{P}}
\def\bbN{\mathbb{N}}
\def\bbE{\mathbb{E}}
\def\bbV{\mathbb{V}}
\def\DS{\mathrm{DS}}

\begin{document}

\title{What is an inductive mean?\\ --- Extended abstract\footnote{This article is an extended version of the paper entitled ``What is... An inductive Mean?''~\cite{InductiveMean-2023} which was published in the Notices of the American Mathematical Society, 70(11):1851-1855 (2023). We kindly acknowledge AMS Notices for allowing us to post this extended abstract on arXiv which contains further references and details.} ---}

\author{Frank Nielsen\\ \ \\ Sony Computer Science Laboratories Inc.\\ Tokyo, Japan}

\date{2024}

\maketitle

\section{Notions of means}
The notion of means~\cite{de2016mean} is central to mathematics and statistics, and  plays a key role in machine learning and data analytics.  
The three classical Pythagorean means of two positive reals $x$ and $y$ are the arithmetic  (A), geometric  (G), 
and  harmonic (H) means, given respectively by
$$
A(x,y) = \frac{x+y}{2},\quad
G(x,y) = \sqrt{xy},\quad
H(x,y) = \frac{2xy}{x+y}.
$$
These  Pythagorean means were originally geometrically studied to define proportions, and the harmonic mean 
led to a beautiful connection between mathematics and music.
The Pythagorean means enjoy the following inequalities:
$$
\min(x,y)\leq H(x,y)\leq G(x,y)\leq A(x,y)\leq \max(x,y),
$$
with equality if and only if $x=y$.
These Pythagorean means belong to a broader parametric family of means, the power means $M_p(x,y)=\left(\frac{x^p+y^p}{2}\right)^\frac{1}{p}$ 
defined for $p\in\bbR\backslash\{0\}$.
We have $A(x,y)=M_1(x,y)$, $H(x,y)=M_{-1}(x,y)$ and in the limits: $G(x,y)=\lim_{p\rightarrow 0} M_p(x,y)$, 
$\max(x,y)=\lim_{p\rightarrow+\infty} M_p(x,y)$ and 
$\min(x,y)=\lim_{p\rightarrow-\infty} M_p(x,y)$.
Power means are also called binomial, Minkowski or H\"older means in the literature.

There are many ways to define and axiomatize means with a rich literature~\cite{bullen2003handbook}.
An important class of means are the quasi-arithmetic means induced by strictly increasing and differentiable real-valued functional generators $f(u)$:
\begin{equation}\label{eq:qam}
M_f(x,y)=f^{-1}\left(\frac{f(x)+f(y)}{2}\right).
\end{equation}
Quasi-arithmetic means satisfy the in-betweenness property of means: 
$\min(x,y)\leq M_f(x,y)\leq \max(x,y)$, and are called so because $f(M_f(x,y))=\frac{f(x)+f(y)}{2}=A(f(x),f(y))$ is the arithmetic mean on the $f$-representation of numbers.

The power means  are quasi-arithmetic means, $M_p=M_{f_p}$, obtained for the following continuous family of generators:
$$
f_p(u)=
\left\{
\begin{array}{ll}
\frac{u^p-1}{p}, & p\in\bbR\backslash\{0\},\cr
\log(u), & p=0.
\end{array}
\right., \quad
f_p^{-1}(u)=
\left\{
\begin{array}{ll}
(1+up)^{\frac{1}{p}}, & p\in\bbR\backslash\{0\},\cr
\exp(u), & p=0.
\end{array}
\right..
$$
Power means are the only homogeneous quasi-arithmetic means, where a mean $M(x,y)$ is said homogeneous when $M(\lambda x,\lambda y)=\lambda\, M(x,y)$ for any $\lambda>0$.   

Quasi-arithmetic means can also be defined for $n$-variable means 
(i.e., $M_f(x_1,\ldots,x_n)=f^{-1}(\frac{1}{n}\sum_{i=1}^n f(x_i))$), and more generally for calculating expected values of random variables~\cite{de2016mean}: 
We denote by $\bbE_f[X]=f^{-1}(\bbE[f(X)])$ 
the quasi-arithmetic expected value of a random variable $X$ induced by a strictly monotone and differentiable function $f(u)$.
For example, the geometric and harmonic expected values of $X$ are defined  by
$\bbE^G[X]=\bbE_{\log x}[X]=\exp(\bbE[\log X])$ and $\bbE^H[X]=\bbE_{x^{-1}}[X]=\frac{1}{\bbE[1/X]}$, respectively.
The ordinary expectation is recovered for $f(u)=u$: $\bbE^A[X]=\bbE_x[X]=\bbE[X]$. 
The quasi-arithmetic expected values satisfy a  strong law of large numbers and a central limit theorem (\cite{de2016mean}, Theorem 1):
Let $X_1,\ldots, X_n$ be independent and identically distributed (i.i.d.) with finite variance $\bbV[f(X)]<\infty$ 
and derivative $f'(\bbE_f[X])\not =0$ at $x=\bbE_f[X]$.
Then we have 
\begin{eqnarray*}
M_f(X_1,\ldots, X_n) &\stackrel{a.s.}{\longrightarrow}& \bbE_f[X]\\
\sqrt{n} \, \left(M_f(X_1,\ldots, X_n)-\bbE_f[X]\right)  &\stackrel{d}{\longrightarrow}& N\left(0, \frac{\bbV[f(X)]}{\left(f'(\bbE_f[X])\right)^2} \right) 
\end{eqnarray*}
as $n\rightarrow\infty$, where $N(\mu,\sigma^2)$ denotes a normal distribution of expectation $\mu$ and variance $\sigma^2$.

\section{Inductive means}
An inductive mean is a mean defined as a limit of a convergence sequence of other means~\cite{sturm2003probability}.
The notion of inductive means defined as limits of sequences was pioneered independently by Lagrange and Gauss~\cite{cox1984arithmetic,borwein1987pi} who studied the following double sequence of iterations: 
\begin{eqnarray*}
a_{t+1} &=& A(a_t,g_t)=\frac{a_t+g_t}{2},\\
g_{t+1} &=& G(a_t,g_t)=\sqrt{a_tg_t},
\end{eqnarray*}
initialized with $a_0=x>0$ and $g_0=y>0$.
We have
$$
g_0\leq \ldots\leq g_t\leq \AGM(x,y)\leq a_t\leq \ldots\leq a_0,
$$
where the homogeneous arithmetic-geometric mean (AGM) is obtained in the limit: 
$$
\AGM(x,y)=\lim_{t\rightarrow\infty} a_t=\lim_{t\rightarrow\infty} g_t.
$$
There is no-closed form formula for the AGM in terms of elementary functions as this induced mean  
 is related to the complete elliptic integral of the first kind $K(\cdot)$~\cite{borwein1987pi}:
$$
\AGM(x,y)=\frac{\pi}{4} \frac{x+y}{K\left(\frac{x-y}{x+y}\right)},
$$
where $K(u)=\int_0^{\frac{\pi}{2}} \frac{\mathrm{d}\theta}{\sqrt{1-u^2\sin^2(\theta)}}$ is the elliptic integral.
The fast quadratic convergence~\cite{ArchimedeanDoubleSequence-1984} of the AGM iterations makes it computationally attractive, and the AGM iterations have been used to numerically calculate digits of $\pi$ or approximate the perimeters of ellipses among others~\cite{borwein1987pi}.

Some inductive means admit closed-form formulas: For example, the arithmetic-harmonic mean $\AHM(x,y)$ obtained as the limit of the double sequence
\begin{eqnarray*}
a_{t+1} &=& A(a_t,h_t)=\frac{a_t+g_t}{2},\\
h_{t+1} &=& H(a_t,h_t)=\frac{2a_th_t}{a_t+h_t},
\end{eqnarray*}
initialized with $a_0=x>0$ and $h_0=y>0$
converges to the geometric mean: 
$$
\AHM(x,y)=\lim_{t\rightarrow\infty} a_t=\lim_{t\rightarrow\infty} h_t=\sqrt{xy}=G(x,y).
$$
In general, inductive means defined as the limits of double sequences with respect to two smooth symmetric means $M_1$ and $M_2$:
\begin{eqnarray*}
a_{t+1} &=& M_1(a_t,b_t),\\
b_{t+1} &=& M_2(a_t,b_t),
\end{eqnarray*}
are proven to convergence quadratically~\cite{ArchimedeanDoubleSequence-1984} to $\DS_{M_1,M_2}(a_0,b_0)=\lim_{t\rightarrow \infty} a_t=\lim_{t\rightarrow \infty} b_t$ (order-$2$ convergence).

\section{Inductive means and matrix means}
We have obtained so far three ways to get the geometric scalar mean $G(x,y)=\sqrt{xy}$ between positive reals $x$ and $y$:
\begin{enumerate}
	\item As an inductive mean with the arithmetic-harmonic double sequence: $G(x,y)=\AHM(x,y)$,
		\item As a quasi-arithmetic mean obtained for the generator $f(u)=\log u$: $G(x,y)=M_{\log}(x,y)$, and
	\item As the limit of power means: $G(x,y)=\lim_{p\rightarrow 0} M_p(x,y)$.
\end{enumerate}

Let us now consider the geometric mean $G(X,Y)$ of two symmetric positive-definite (SPD) matrices $X$ and $Y$ of size $d\times d$.
SPD matrices generalize positive reals.
We shall investigate the three generalizations of the above approaches of the scalar geometric mean, and show that they yield different notions of matrix geometric means when $d>1$. 

\subsection{First generalization for the matrix geometric mean}
First, the AHM iterations can be extended to SPD matrices instead of reals:
\begin{eqnarray*}
A_{t+1} &=& \frac{A_t+H_t}{2}=A(A_t,H_t),\\
H_{t+1} &=& 2\, (A_t^{-1}+H_t^{-1})^{-1}=H(A_t,H_t),
\end{eqnarray*}
where the matrix arithmetic mean is $A(X,Y)=\frac{X+Y}{2}$ and the matrix harmonic mean is $H(X,Y)=2(X^{-1}+Y^{-1})^{-1}$.
The AHM iterations initialized with $A_0=X$ and $H_0=Y$ yield in the limit $t \rightarrow\infty$, the matrix arithmetic-harmonic mean~\cite{AHM-Nakamura-2001,ConvexFunctionalMean-2001} (AHM):
$$
\AHM(X,Y)=\lim_{t\rightarrow +\infty} A_t=\lim_{t\rightarrow +\infty} H_t.
$$
Remarkably, the matrix AHM enjoys quadratic convergence to the following SPD matrix:
$$
\AHM(X,Y)=X^{\frac{1}{2}}\, (X^{-\frac{1}{2}}\, Y\, X^{-\frac{1}{2}})^{\frac{1}{2}}\, X^{\frac{1}{2}}=G(X,Y).
$$
When $X=x$ and $Y=y$ are positive reals, we recover $G(X,Y)=\sqrt{xy}$.
When $X=I$, the identity matrix, we get $G(I,Y)=Y^{\frac{1}{2}}=\sqrt{Y}$, the positive square root of SPD matrix $Y$~\cite{MatrixSquareRoot-Sra-2016}.
Thus the matrix AHM iterations provide a fast method in practice to numerically approximate matrix square roots by bypassing the matrix eigendecomposition.
When matrices $X$ and $Y$ commute (i.e., $XY=YX$), we have $G(X,Y)=\sqrt{XY}$.
The geometric mean $G(A,B)$ is proven to be the unique solution to the matrix Ricatti equation $XA^{-1}X=B$,
is invariant under inversion (i.e., $G(A,B)=G(A^{-1},B^{-1})^{-1}$), and satisfies the determinant property $\det(G(A,B))=\sqrt{\det(A)\,\det(B)}$.

Let $\bbP$ denote the set of symmetric positive-definite $d\times d$ matrices.  
The matrix geometric mean can  be interpreted using a Riemannian geometry~\cite{moakher2005differential,bhatia2006riemannian} of the cone $\bbP$:
Equip $\bbP$ with the trace metric tensor, i.e., a collection of smoothly varying inner products $g_P$ for $P\in\bbP$ defined by
$$
g_P(S_1,S_2)=\tr\left(P^{-1} S_1 P^{-1} S_2\right),
$$ 
where $S_1$ and $S_2$ are matrices belonging to the vector space of symmetric $d\times d$ matrices (i.e., $S_1$ and $S_2$ are geometrically vectors  of the tangent plane $T_P$ of $P\in\bbP$).
The geodesic length distance on the Riemannian manifold $(\bbP,g)$ is
$$
\rho(P_1,P_2)=\left\| \log\left(P_1^{-\frac{1}{2}}\, P_2\, P_1^{-\frac{1}{2}}\right)\right\|_F  =\sqrt{\sum_{i=1}^d \log^2 \lambda_i\left(P_1^{-\frac{1}{2}}\, P_2\, P_1^{-\frac{1}{2}}\right)},
$$
where $\lambda_i(M)$ denotes the $i$-th largest real eigenvalue of a symmetric matrix $M$, $\|\cdot\|_F$ denotes the Frobenius norm, and $\log P$ is the unique matrix logarithm of a SPD matrix $P$.
Interestingly, the matrix geometric mean $G(X,Y)=\AHM(X,Y)$ can also  be interpreted as the Riemannian center of mass of $X$ and $Y$:
$$
G(X,Y)=\arg\min_{P\in\bbP} \frac{1}{2}\rho^2(X,P)+\frac{1}{2}\rho^2(Y,P).
$$
This  Riemannian least squares mean is also called the Cartan, K\"archer, or Fr\'echet mean~\cite{RecursiveFrechetMeanNPC-2016} in the literature.
More generally, the Riemannian geodesic $\gamma(X,Y;t)=X\#_t Y$ between $X$ and $Y$ of $(\bbP,g)$ for $t\in [0,1]$ is expressed using the weighted matrix geometric mean $G(X,Y;1-t,t)=X\#_t Y$ minimizing
$$
(1-t) \rho^2(X,P)+t\rho^2(Y,P).
$$
This Riemannian barycenter can be solved as
$$
X\#_t Y= 
X^{\frac{1}{2}}\, \left(X^{-\frac{1}{2}}\, Y\, X^{-\frac{1}{2}}\right)^t\, X^{\frac{1}{2}},
$$
with $G(X,Y)=X\#_{\frac{1}{2}} Y$, $X\#_t Y=Y\#_{1-t} X$, and $\rho(X\#_t Y,X)=t\, \rho(X,Y)$, i.e., $t$ is the arc length parameterization of the constant speed geodesic $\gamma(X,Y;t)$. When matrices $X$ and $Y$ commute, we have $X\#_t Y=X^{1-t}Y^t$.
We thus interpret the matrix geometric mean $G(X,Y)=X\#Y=X\#_{\frac{1}{2}} Y$ as the Riemannian geodesic midpoint.

The matrix geometric mean can also be interpreted as the centroid of the $S$-divergence~\cite{Sra-SDivergence-2016}:
$$
G(X,Y)=\arg\min_{C\in\bbP} S_\ld(C,X)+S_\ld(C,Y),
$$
where $S_\ld$ is the $S$-divergence, a symmetrization of the log-det divergence (also called Stein divergence in statistics):
$$
S_\ld(X,Y)=\log\det\left(\frac{X+Y}{2}\right)-\frac{1}{2}\log\det(XY).
$$

In general, the Gauss double sequence has been generalized to inductive $(A,M_{\nabla F})$-mean~\cite{nielsen-2024} where $A(\theta_1,\theta_2)=\frac{\theta_1+\theta_2}{2}$ is the arithmetic mean and 
$M_{\nabla F}(\theta_1,\theta_2)=(\nabla F)^{-1}\left(\frac{\nabla F(\theta_1)+\nabla F(\theta_2)}{2}\right)$ is a quasi-arithmetic center (i.e., multivariate quasi-arithmetic mean) induced by the gradient of a Legendre type function~\cite{LegendreType-1967} $F(\theta)$. 
For a scalar weight $w\in(0,1)$, let $M_{\nabla F}(\theta_1,\theta_2;w,1-w)=(\nabla F)^{-1}\left(w\nabla F(\theta_1)+(1-w)\nabla F(\theta_2)\right)$.

\subsection{Second generalization for the matrix geometric mean}
Second, let us consider the matrix geometric mean as the limit of  matrix quasi-arithmetic power means which can be  defined~\cite{lim2012matrix,audenaert2013matrix} as $Q_p(X,Y)=(X^p+Y^p)^{\frac{1}{p}}$ for $p\in\bbR, p\not=0$, with 
$Q_1(X,Y)=A(X,Y)$ and $Q_{-1}(X,Y)=H(X,Y)$.
We get $\lim_{p\rightarrow 0} Q_p(X,Y)=\LEM(X,Y)$, the log-Euclidean matrix mean defined by
$$
\LEM(X,Y)=\exp\left(\frac{\log X+\log Y}{2}\right),
$$
where $\exp$ and $\log$ denote the matrix exponential and  the matrix logarithm, respectively.
We have $\LEM(X,Y)\not =G(X,Y)$.
Consider the Loewner partial order $\preceq$ on the cone $\bbP$:
 $P\preceq Q$ if and only if $Q-P$ is positive semi-definite.
A mean $M(X,Y)$ is said operator monotone~\cite{bhatia2006riemannian} if for $X'\preceq X$ and $Y'\preceq Y$, we have
$M(X',Y')\preceq M(X,Y)$. 
The log-Euclidean mean $\LEM(X,Y)$ is not operator monotone but the Riemannian geometric matrix mean $G(X,Y)$ is operator monotone.

\subsection{Third generalization for the matrix geometric mean}
Third, we can define matrix power means $M_p(X,Y)$ for $p\in (0,1]$  
by uniquely solving the following matrix equation~\cite{lim2012matrix}:
\begin{equation}\label{eq:mp}
M=\frac{1}{2} M\#_p X + \frac{1}{2}  M\#_p Y.
\end{equation}
Let $M_p(X,Y)=M$ denote the unique solution of Eq.~\ref{eq:mp}.
This equation is the matrix analogue of the scalar equation $m=\frac{1}{2} m^{1-p}x^p + \frac{1}{2} m^{1-p}y^p$  which can be solved as 
$m=\left(\frac{1}{2}x^p+\frac{1}{2}y^p\right)^{\frac{1}{p}}=M_p(x,y)$, i.e., the scalar $p$-power mean.
In the limit case $p\rightarrow 0$, this matrix power mean $M_p$ yields the matrix geometric/Riemannian mean~\cite{lim2012matrix}: 
$$
\lim_{p\rightarrow 0^+} M_p(X,Y)=G(X,Y). 
$$
In general, we get the following closed-form expression~\cite{lim2012matrix} of  this matrix power mean for $p\in (0,1)$:
$$
M_p(X,Y)= X \#_{\frac{1}{p}}\left (\frac{1}{2}X+ \frac{1}{2}(X \#_p Y)\right).
$$

\section{Inductive means, circumcenters, and medians of several matrices}

To extend these various binary matrix means of two matrices  to matrix means of $n$ matrices $P_1,\ldots, P_n$ of $\bbP$, we can use induction sequences~\cite{RecursiveFrechetMeanNPC-2016}.
First, the $n$-variable matrix geometric mean $G(P_1,\ldots, P_n)$ can be defined as the unique Riemannian center of mass:
$$
G(P_1,\ldots, P_n)= \arg\min_{P\in\bbP} \sum_{i=1}^n \frac{1}{n}\rho^2(P,P_i).
$$
This geometric matrix mean $G=G(P_1,\ldots, P_n)$ can be characterized as the unique solution of 
$\sum_{i=1}^n \log\left(G^{-\frac{1}{2}}P_iG^{-\frac{1}{2}}\right)=0$ (called the K\"archer equation), and is proven to satisfy the ten Ando-Li-Mathias properties~\cite{ALM-2004} defining what should be a good matrix generalization of the scalar geometric mean.

Holbrook~\cite{holbrook2012no} proposed the following sequence of iterations to approximate  $G(P_1,\ldots, P_n)$:
\begin{equation}\label{eq:igm}
M_{t+1}=M_t \#_{\frac{1}{t+1}} P_{t\ \mod\ n}
\end{equation}
with $M_1$ initialized to $P_1$.
In the limit $t\rightarrow\infty$, we get the $n$-variable geometric mean: $\lim_{t\rightarrow\infty} M_t=G(P_1,\ldots,P_n)$.
This deterministic inductive definition of the matrix geometric mean by Eq.~\ref{eq:igm} allows to prove that 
the geometric mean $G(P_1,\ldots,P_n)$ is monotone~\cite{holbrook2012no}:
That is, if $P_1'\preceq P_1$, \ldots, $P_n'\preceq P_n$ then we have
$G(P_1',\ldots,P_n')\preceq G(P_1,\ldots,P_n)$.
The following matrix  arithmetic-geometric-harmonic mean inequalities extends the scalar case: 
$$
H(X,Y;1-t,t)=((1-t)X^{-1}+tY^{-1})^{-1} \preceq G(X,Y;1-t,t) \preceq A(X,Y;1-t,t)=(1-t)X+tY.
$$

Now, if instead of taking cyclically the input matrices $P_1,\ldots, P_n, P_1,\ldots,P_n,\ldots$, we choose at iteration $t$ the farthest matrix in $P_1,\ldots, P_n$ to $M_t$ with respect to the Riemannian distance $\rho$, we get the Riemannian circumcenter~\cite{arnaudon2013approximating} $C(P_1,\ldots,P_n)$ which is the minimax minimizer: 
$$
C(P_1,\ldots,P_n)=\arg\min_{C\in\bbP} \max_{i\in\{1,\ldots,n\}} \rho(P_i,C).
$$
The sequence of iterations  
\begin{equation}\label{eq:minimax}
C_{t+1}=C_t \#_{\frac{1}{t+1}} P_{\farthest(t)},
\end{equation}
where
$$
\farthest(t)=\arg\max_{i\in\{1,\ldots,n\}} \rho(C_t,P_i),
$$
 initialized with $C_1=P_1$ is such that
$$
C(P_1,\ldots,P_n)=\lim_{t\rightarrow\infty} C_t.
$$
The uniqueness of the smallest enclosing ball and the proof of convergence of the iterations of Eq.~\ref{eq:minimax}  relies on the fact that the cone $\bbP$ is of non-positive sectional curvatures~\cite{arnaudon2013approximating}:
$\bbP$ is a Non-Positive Curvature space or NPC space for short.
Note that the convergence of the inductive Riemannian barycenter based on improved shuffle sequences instead of the cyclic sequence has been proven for arbitrary non-positive curvature (NPC) manifolds in~\cite{massart2018matrix}.

The inductive matrix geometric mean $G(X,Y)=\AHM(X,Y)$ can be interpreted geometrically~\cite{AHM-Nakamura-2001} by considering the manifold 
$(\bbP,g,\nabla,\nabla^*)$ in information geometry where $\nabla$ is the torsion-free affine connection defining the mixture geodesic 
$\gamma^\nabla(P_1,P_2;t)=(1-t)P_1+t P_2$ (weighted matrix arithmetic mean) and $\nabla^*$ is the dual torsion-free affine connection defining the dual geodesic
$\gamma^{\nabla^*}(P_1,P_2;t)=((1-t)P_1^{-1}+t P_2^{-1})^{-1}$ (weighted matrix harmonic mean).
The connections $\nabla$ and $\nabla^*$ are called dual in information geometry~\cite{DFS-SPD-1996} because they are coupled to the Riemannian metric $g$ in a sense that the mid-connection $\bar\nabla=\frac{\nabla+\nabla^*}{2}$ yields the Levi-Civita connection induced by the metric tensor $g$ with corresponding Riemannian geodesic
$\gamma^{\bar\nabla}(P_1,P_2;t)=P_1\#_t P_2$.
A connection  $\nabla$  is flat when the geodesic $\gamma^\nabla$ can be expressed in a coordinate system $\theta$ as linear interpolation:
$$
\theta(\gamma^\nabla(P_1,P_2;t))=(1-t)\theta(P_1)+t \theta(P_2).
$$ 
The coordinate system $\theta$ is then  called an affine coordinate system of $\nabla$~\cite{DFS-SPD-1996}.
The manifold $(\bbP,g,\nabla,\nabla^*)$ is a dually flat but Riemannian negatively curved manifold $(\PD(d),g)$, with dual affine coordinate systems $\theta(P)=P$ and $\theta^*(P)=P^{-1}$.
In general, in a dually flat manifold $(M,g,\nabla,\nabla^*)$, there exists dual potential functions $F(\theta)$ and $F^*(\eta)$ related by the Legendre-Fenchel transform: 
$F^*(\eta)=\sup_{\theta\in\Theta} \{\inner{\theta}{\eta}-F(\theta)\}$ (where $\inner{\cdot}{\cdot}$ denotes an inner product like the vector dot product of matrix Hilbert-Schmidt inner product) with $\eta=\nabla F(\theta)$ and $\theta=\nabla F^*(\eta)$ where $\nabla F=(\nabla F^*)^{-1}$ and $\nabla F^*=(\nabla F)^{-1}$ denote the reciprocal gradient  vectors. Dually flat manifolds are Hessian manifolds~\cite{shima2007geometry} because the metric $g$ can be expressed in the $\theta$- or $\eta$-coordinate system by the Hessians: $g(\theta)=\nabla^2 F(\theta)=\nabla\eta(\theta)$ and $g(\eta)=\nabla^2 F^*(\eta)=\nabla \theta(\eta)$.
The corresponding dual geodesics $\gamma(p_1,p_2;t)$ and $\gamma^*(p_1,p_2;t)$ between points $p_1$ and $p_2$ of a dually flat manifold $(M,g,\nabla,\nabla^*)$ with corresponding dual coordinates $\theta(p_1)=\theta_1$, $\eta(p_1)=\eta_1$ and $\theta(p_2)=\theta_2$, $\eta(p_2)=\eta_2$ can be expressed using generalized quasi-arithmetic means as follows:
\begin{eqnarray*}
\theta(\gamma(p_1,p_2;t)) &=& (1-t)\theta(p_1)+t\theta(p_2),\\
 \eta(\gamma(p_1,p_2;t)) &=& \nabla F(\gamma(p_1,p_2;t)) = (\nabla F^*)^{-1}\left( (1-t)\nabla F^*(\eta_1)+t\nabla F^*(\eta_2)\right),\\
&=:& M_{\nabla F^*}(\eta_1,\eta_2;1-t,1).
\end{eqnarray*}
Similarly, we have $\eta(\gamma^*(p_1,p_2;t))=M_{\nabla F}(\theta_1,\theta_2;1-t,t)$.
Note that in general multivariate functions are not globally invertible but only locally invertible via the implicit function theorem~\cite{krantz2002implicit}.
However,  the gradient functions $\nabla F(\theta)$ of Legendre-type functions $F(\theta)$ are always globally invertible, and allows to  neatly generalize
the  quasi-arithmetic means of Eq.~\ref{eq:qam} as follows:
\begin{eqnarray*}
M_{\nabla F}(\theta_1,\theta_2)&=&\nabla F^{-1}\left(\frac{\nabla F(\theta_1)+\nabla F(\theta_2)}{2}\right),\\
&=& \nabla F^*\left(\frac{\nabla F(\theta_1)+\nabla F(\theta_2)}{2}\right),
\end{eqnarray*}
where $F^*(\eta)$ denote the Legendre-Fenchel convex conjugate of $F(\theta)$.

The Riemannian median minimizing $\arg\min_{P\in\bbP} \sum_{i=1}^n \frac{1}{n}\rho(P,P_i)$
is proven to be unique in Riemannian NPC spaces, and can be obtained as the limit of the following cyclic order sequence~\cite{bacak2014computing}:
\begin{eqnarray*}
X_{kn+1} &=& X_{kn}\#_{t_{k,1}} P_1,\\
X_{kn+2} &=& X_{kn+1}\#_{t_{k,2}} P_2,\\
\vdots &=& \vdots \\
X_{kn+n} &=& X_{kn+n-1}\#_{t_{k,n}} P_n,
\end{eqnarray*}
where $t_{k,n}=\min\left\{1,\frac{\lambda_k}{n\, \rho(P_n,X_{kn+n-1})}\right\}$ with the positive real sequence $(\lambda_k)$ such that $\sum_{k=0}^\infty \lambda_k=\infty$ and $\sum_{k=0}^\infty \lambda_k^2<\infty$ (e.g., $\lambda_k=\frac{1}{k+1}$).

Finally, let us mention that 
Bini, Meini, and Poloni~\cite{bini2010effective} proposed a class of recursive geometric matrix means 
$G_{s_1,\ldots, s_{n-1}}(P_1,\ldots, P_n)$ parameterized by $(n-1)$-tuple of scalar parameters, and defined recursively as the common limit of the following sequences:
$$
P_i^{(r+1)}= P_i^{(r)}\#_{s_1} G_{s_2,\ldots, s_{n-1}}\left(P_1,\ldots,P_{i-1},P_{i+1},\ldots,P_n\right),\quad i\in\{1,\ldots, n\}.
$$
In particular, these matrix means exhibit a unique $(n-1)$-tuple for which the recursive mean 
$G_{\frac{n-1}{n},\frac{n-2}{n-1},\ldots,\frac{1}{2}}(P_1,\ldots, P_n)$ converges fast in cubic order (order-$3$ convergence). 
This geometric matrix mean is called the BMP mean in the literature. 
Furthermore, the mean $G_{1,1,\ldots,1,\frac{1}{2}}(P_1,\ldots, P_n)$ coincides  with the Ando-Li-Mathias geometric mean~\cite{ALM-2004} (ALM) which convergences linearly.

\section{Random variables, expectations, and the law of large numbers}

Although inductive means as limits of sequences have been considered since the 18th Century (AGM by Lagrange and Gauss), 
this term was only recently coined by Karl-Theodor Sturm in 2003 (see Definition 4.6 in \cite{sturm2003probability}) who considered inductive sequences to calculate probability expectations of random variables on non-positive curvature complete metric spaces.
For example, let $\calP(\bbP)$ denote the set of probability measures on $\bbP$ with bounded support~\cite{sturm2003probability}.
Let $X:\Omega\rightarrow\bbP$ be a SPD-valued random variable with probability density function $p_X$ expressed with respect to the 
canonical Riemannian volume measure $\domega(P)=\sqrt{\det(g_P)}$.
The expectation $\bbE[X]$ and the variance $\bbV[X]$ of a random variable $X\sim p_X$ are defined respectively as 
the unique minimizer of $C\mapsto \bbE[\rho^2(X,C)]=\int_\bbP \rho^2(C,P)p_X(P)\domega(P)$ and $\inf_{P\in\bbP} \bbE[\rho^2(X,P)]$.
Consider $(X_i)_{i\in\bbN}$ to be an independent sequence of measurable maps $X_i:\Omega\rightarrow\bbP$ with identical distributions $p_{X_i}=p_X$, and let $p_n=\frac{1}{n}\sum_{i=1}^n \delta_{X_i}\in\calP(\bbP)$ denote the empirical distribution. 
Then the following empirical law of large numbers  holds as $n\rightarrow\infty$:
$$
G(X_1,\ldots,X_n)\rightarrow \bbE[X].
$$
Several proofs are reported in the literature (e.g., Proposition 6.6 of~\cite{sturm2003probability}, Theorem~1 of~\cite{RecursiveFrechetMeanNPC-2016}, or Theorem 5.1 of~\cite{bacak2014computing}).
Thus the expectation $\bbE[X]$ of a SPD-valued random variable can be estimated incrementally by 
considering increasing sequences $(X_i)_{i\in\bbN}$ of i.i.d. random vectors, and incrementally computing their Riemannian means.
Experiments demonstrating convergence to various probability law expectations $p_X$ are reported in~\cite{RecursiveFrechetMeanNPC-2016}.

\section{Closing remarks}
The AHM double sequence yielding the matrix geometric mean can further be generalized to define self-dual operators on convex functionals in Hilbert spaces~\cite{ConvexFunctionalMean-2001} based on 
the Legendre-Fenchel transformation  (called convex geometric mean functionals).
For example, the AHM iterations initialized on a pair of non-zero complex numbers 
$z_1=r_1\, e^{i\theta_1}$ and $z_2=r_2\, e^{i\theta_2}$ expressed in polar forms is proven to converge to $\AHM(z_1,z_2)=\sqrt{r_1r_2}\, e^{i\frac{\theta_1+\theta_2}{2}}$ which involves both the scalar arithmetic mean $A(\theta_1,\theta_2)$ and the scalar geometric mean $G(r_1,r_2)$.

To conclude, let us say that not only  is it important to consider which mean we mean~\cite{de2016mean} 
but it is also essential to state which matrix geometric mean we mean!

\bibliographystyle{plain}
\bibliography{WhatIsInductiveMeanArXivBIB}
\end{document}